\documentclass[amsmath,aps,prd,showpacs,a4paper,10pt]{revtex4}

\begin{document}

\newcommand{\be}[1]{\begin{equation}\label{#1}}
\newcommand{\ee}{\end{equation}}
\newcommand{\bea}{\begin{eqnarray}}
\newcommand{\eea}{\end{eqnarray}}
\def\disp{\displaystyle}

\begin{titlepage}

\begin{flushright}
astro-ph/0603052
\end{flushright}

\title{\Large \bf Interacting Vector-like Dark Energy, the First and Second
Cosmological Coincidence Problems}

\author{Hao Wei}
\email{haowei@itp.ac.cn}
\affiliation{Institute of Theoretical Physics, Chinese Academy of Sciences,
P.O. Box 2735, Beijing 100080, China \\
Graduate School of the Chinese Academy of Sciences, Beijing 100039, China}

\author{Rong-Gen Cai}
\email{cairg@itp.ac.cn}
\affiliation{Institute of Theoretical Physics, Chinese Academy of Sciences,
P.O. Box 2735, Beijing 100080, China \\
Department of Physics, Osaka University, Toyonaka, Osaka 560-0043, Japan}

\begin{abstract}
\vspace{5mm}
\centerline{\bf ABSTRACT}\vspace{2mm}
One of the puzzles of the dark energy problem is the (first) cosmological 
coincidence problem, namely, {\em why does our universe begin the accelerated 
expansion recently? why are we living in an epoch in which the dark energy 
density and the dust matter energy density are comparable?} On the other hand, 
cosmological observations hint that the equation-of-state parameter (EoS) of dark 
energy crossed the phantom divide $w_{de}=-1$ in the near past. Many dark 
energy models whose EoS can cross the phantom divide have been proposed. 
However, to our knowledge, these models with crossing the phantom divide only 
provide the possibility that $w_{de}$ can cross $-1$. They do not answer another 
question, namely, {\em why crossing the phantom divide occurs recently?} Since 
in many existing models whose EoS can cross the phantom divide, $w_{de}$ 
undulates around $-1$ randomly, {\em why are we living in an epoch 
$w_{de}<-1$?} This can be regarded as the second cosmological coincidence 
problem. In this work, the cosmological evolution of the vector-like dark energy 
interacting with background perfect fluid is investigated. We find that the first 
and second cosmological coincidence problems can be alleviated at the same time 
in this scenario.
\end{abstract}

\pacs{95.36.+x, 98.80.Cq, 98.80.-k, 45.30.+s}

\maketitle

\end{titlepage}

\renewcommand{\baselinestretch}{1.6}



\section{\label{sec1} Introduction}

Dark energy problem~\cite{r1} has been one of the most active 
fields in modern cosmology, since the discovery of accelerated
expansion of our universe~\cite{r2,r3,r4,r5,r6,r61}. In the
observational cosmology, the equation-of-state parameter (EoS) of
dark energy $w_{de}\equiv p_{de}/\rho_{de}$ plays a central role,
where $p_{de}$ and $\rho_{de}$ are its pressure and energy
density, respectively. To accelerate the expansion, the EoS of
dark energy must satisfy $w_{de}<-1/3$. The simplest candidate of
the dark energy is a tiny positive time-independent cosmological
constant $\Lambda$, whose EoS is $-1$. However, it is difficult to
understand why the cosmological constant is about 120 orders of
magnitude smaller than its natural expectation, i.e. the Planck
energy density. This is the so-called cosmological constant
problem. Another puzzle of the dark energy is the (first)
cosmological coincidence problem, namely, {\em why does our
universe begin the accelerated expansion recently? why are we
living in an epoch in which the dark energy density and the dust
matter energy density are comparable?} This problem becomes very
serious especially for the cosmological constant as the dark
energy candidate. The cosmological constant remains unchanged
while the energy densities of dust matter and radiation decrease
rapidly with the expansion of our universe. Thus, it is necessary
to make some fine-tunings. In order to give a reasonable
interpretation to the (first) cosmological coincidence problem,
many dynamical dark energy models have been proposed as
alternatives to the cosmological constant, such as
quintessence~\cite{r7,r8}, phantom~\cite{r9,r10,r11},
k-essence~\cite{r12} etc.

Recently, by fitting the SNe Ia data, marginal evidence for
$w_{de}(z)<-1$ at redshift $z<0.2$ has been found~\cite{r13}. In
addition, many best-fits of the present value of $w_{de}$ are less
than $-1$ in various data fittings with different
parameterizations (see~\cite{r14} for a recent review). The
present observational data seem to slightly favor an evolving dark
energy with $w_{de}$ crossing $-1$  from above to below in the
near past~\cite{r15}. Obviously, the EoS of dark energy $w_{de}$
cannot cross the so-called phantom divide $w_{de}=-1$ for
quintessence or phantom alone. Although it seems possible for some
variants of k-essence to give a promising solution to cross the
phantom divide, a no-go theorem, shown in~\cite{r16}, shatters
this kind of hopes. In fact, it is not a trivial task to build
dark energy model whose EoS can cross the phantom divide. To this
end, a lot of efforts~\cite{r17,r18,r19,r20,r21,r22,r23,r24,
r25,r26,r27,r28,r29,r30,r31,r32,r59} have been made. However, to our
knowledge, many of these models only provide the possibility that
$w_{de}$ can cross $-1$. They do not answer another question,
namely, {\em why crossing phantom divide occurs recently?} Since
in many existing models whose EoS can cross the phantom divide,
$w_{de}$ undulates around $-1$ randomly, {\em why are we living in
an epoch $w_{de}<-1$?} It can be regarded as the second cosmological 
coincidence problem~\cite{r33}.

The so-called second cosmological coincidence problem was
seriously discussed in~\cite{r33} for the first time.
In~\cite{r33}, the key point is the trigger mechanism, similar to
the case of hybrid inflation~\cite{r34} and the model by Gong and
Kim~\cite{r35}. In the hybrid dark energy model considered
in~\cite{r33}, a quintessence and a phantom are employed. The
feature of spontaneous symmetry breaking plays a critical role. In
the first stage, the phantom field $\sigma$ is trapped to
$\sigma=0$, while the quintessence field $\phi$ slowly rolls down
for a long time. In this stage, the effective EoS of hybrid dark
energy remains larger than $-1$. When $\phi$ reaches a critical
value $\phi_c$, a phase transition takes place and the phantom
field $\sigma$ is triggered to climb up its potential. And then,
the quintessence field $\phi$ is trapped to $\phi=0$. Crossing the
phantom divide occurs between the moment of phase transition
$\phi=\phi_c$ and the moment of $\phi$ being trapped at $\phi=0$
eventually. The effective EoS remains smaller than $-1$ in the
major part of the second stage. When $\sigma$ continuously climbs
up, it will eventually arrive at a critical value $\sigma_c$. The
quintessence field $\phi$ is triggered to roll down again. In the
third stage, whether the EoS remains smaller than $-1$ or changes
to be larger than $-1$ depends on the profile of the effective
potential of hybrid dark energy in this stage. Thus, the avoidance
of the big rip is possible for suitable model parameters. We refer 
to the original paper~\cite{r33} for more details. (Also, see~\cite{r60} 
for other discussion on the second cosmological coincidence problem, 
soon after our previous paper~\cite{r33}.)

Although the scalar field is used extensively because of its
simplicity, the vector field also gets its applications in modern
cosmology. Here, we only mention a few, such as in~\cite{r36,r37}, the
vector field is considered as the source which drives the
inflation; in~\cite{r38,r39} the Lorentz-violated vector field and
its effects to universe are studied; the quantum fluctuations of
vector fields produced at the first stage of reheating after
inflation is also studied in \cite{r40}; the cosmology of massive
vector fields with SO(3) global symmetry is investigated
in~\cite{r41}; and also see \cite{r42,r43} and references therein
for the literature on the magnetic fields in the universe. It is
found that non-linear electromagnetic field can drive the
acceleration of the universe~\cite{r44}. The vector field can be a
good dark matter candidate reproducing flat rotation curves in
dark halos of spiral galaxies (see~\cite{r45} for example). Of
course, the vector field is also a viable dark energy
candidate~\cite{r46,r47}. The effects of vector field dark energy candidate
on the cosmic microwave background radiation and the large scale structure
are discussed in~\cite{r58}.

In this work, the cosmological evolution of the vector-like dark
energy proposed in~\cite{r46} interacting with background perfect
fluid is investigated. In fact, in~\cite{r46} the cosmological
evolution of the vector-like dark energy with inverse power-law
potential and {\em without} interaction with background matter was
studied, by means of directly numerical and approximate solutions
of the equation of motion in two limits of matter domination and
vector-like dark energy domination. Differing from~\cite{r46}, we
investigate the cosmological evolution of the vector-like dark
energy and background perfect fluid in this work by means of
dynamical system~\cite{r48}. And we study the models with not only
inverse power-law potential but also exponential potential.
Furthermore, we consider the cases of vector-like dark energy
interacting with background perfect fluid, while the interaction
terms are taken to be four different forms which are familiar in
the literature. We find that the first and second cosmological
coincidence problems can be alleviated at the same time in this
scenario.

This paper is organized as follows. In Sec.~\ref{sec2}, we will
briefly present the main points of the vector-like dark energy
model proposed in~\cite{r46}. In Sec.~\ref{sec3}, we give out the
equations of the dynamical system of vector-like dark energy with
interaction to background perfect fluid for the most general case.
That is, we leave the potential of vector-like dark energy and the
interaction form undetermined. We will investigate the dynamical
system for the models with inverse power-law and exponential
potentials in Sec.~\ref{sec4} and~\ref{sec5}, respectively. In
each case with different potential, we consider four different
interaction forms between vector-like dark energy and background
perfect fluid. The interaction forms are taken to be the most
familiar interaction ones considered in the literature. In
Sec.~\ref{sec6}, the first and second cosmological coincidence
problems are discussed. Finally, brief conclusion and discussion
are given in Sec.~\ref{sec7}.

We use the units $\hbar=c=1$ and $\kappa^2\equiv 8\pi G$ throughout this paper.


\section{\label{sec2} Vector-like dark energy}
In fact, the ``vector-like dark energy'' is a so-called ``cosmic
triad'' (see~\cite{r46}), which is a set of three identical
vectors pointing in mutually orthogonal directions, in order to
avoid violations of isotropy. Following~\cite{r46}, we consider
the case of vector-like dark energy minimally coupled to gravity,
and the action is given by \be{eq1} S=\int d^4
x\,\sqrt{-g}\left\{\frac{R}{2\kappa^2}-\sum\limits_{a=1}^3\left[\frac{1}{4}F_{\mu\nu}^a
F^{a\,\mu\nu}+V\left(A^{a2}\right)\right]+{\cal L}_m
\left(g_{\mu\nu},\psi\right)\right\}, \ee where $g$ is the
determinant of the metric $g_{\mu\nu}$, $R$ is the Ricci scalar,
$F_{\mu\nu}^a\equiv\partial_\mu A_\nu^a-\partial_\nu A_\mu^a$,
$A^{a2}\equiv g^{\mu\nu}A_\mu^a A_\nu^a$, and ${\cal L}_m$ is the
Lagrangian density of matter fields $\psi$. Latin indices label
the different fields ($a,b,\ldots=1,2,3$) and Greek indices label
different spacetime components ($\mu,\nu,\ldots=0,1,2,3$).
Actually, the number of vector fields is dictated by the number of
spatial dimensions and the requirement of isotropy~\cite{r46}. The
Latin indices are raised and lowered with the flat ``metric''
$\delta_{ab}$. The potential term $V\left(A^{a2}\right)$
explicitly violates gauge invariance. In~\cite{r46}, it is argued
that this kind of ``cosmic triad'' can naturally arise from gauge
theory with a single $SU(2)$ gauge group.

From action (\ref{eq1}), one can get the energy-momentum tensor of the cosmic triad
and the equations of motion for the vectors $A_\mu^a$ as
\be{eq2}
^{(A)}T_{\mu\nu}=\sum\limits_a\left\{F_{\mu\rho}^a F_\nu^{a\,\rho}+
2\frac{dV}{dA^{a2}}A_\mu^a A_\nu^a-\left[\frac{1}{4}F_{\rho\sigma}^a F^{a\,\rho\sigma}+
V\left(A^{a2}\right)\right]g_{\mu\nu}\right\},
\ee
and
\be{eq3}
\partial_\mu\left(\sqrt{-g}F^{a\,\mu\nu}\right)=2\sqrt{-g}\frac{dV}{dA^{a2}}A^{a\,\nu},
\ee respectively. We consider a spatially flat Friedmann-Robertson-Walker (FRW) 
universe with metric
\be{eq4} ds^2=-dt^2+a^2(t)d{\bf x}^2, \ee where $a$ is the scale
factor. An ansatz for the vectors, which can be compatible with
homogeneity and isotropy, is \be{eq5} A^b_{\ \mu}=\delta^b_{\
\mu}A(t)\cdot a. \ee Thus, the three vectors point in mutually
orthogonal spatial directions, and share the same time-dependent
length, i.e. $A^{a2}\equiv A_\mu^a A^{a\,\mu}=A^2(t)$.
Substituting Eqs.~(\ref{eq5}) and (\ref{eq4}) into
Eq.~(\ref{eq3}), one obtains \be{eq6}
\ddot{A}+3H\dot{A}+\left(H^2+\frac{\ddot{a}}{a}\right)A+\frac{dV}{dA}=0,
\ee where $H\equiv\dot{a}/a$ is the Hubble parameter, and a dot
denotes the derivative with respect to the cosmic time $t$. The
Friedmann equation and Raychaudhuri equation are given by,
respectively, \be{eq7}
H^2=\frac{\kappa^2}{3}\rho_{tot}=\frac{\kappa^2}{3}\left(\rho_A+\rho_m\right),
\ee and \be{eq8}
\dot{H}=-\frac{\kappa^2}{2}\left(\rho_{tot}+p_{tot}\right)
=-\frac{\kappa^2}{2}\left(\rho_A+\rho_m+p_A+p_m\right), \ee where
$p_m$ and $\rho_m$ are the pressure and energy density of
background matter, respectively. The energy density and
(isotropic) pressure of the vector-like dark energy are given by
\bea
&&\rho_A=\frac{3}{2}\left(\dot{A}+HA\right)^2+3V\left(A^2\right),\label{eq9}\\
&&p_A=\frac{1}{2}\left(\dot{A}+HA\right)^2-3V\left(A^2\right)+2\frac{dV}{dA^2}A^2,\label{eq10}
\eea
respectively. Noting that $\ddot{a}/a=H^2+\dot{H}$, one can check that Eq.~(\ref{eq6}) is
equivalent to the energy conservation equation of vector-like dark energy, namely
$\dot{\rho}_A+3H(\rho_A+p_A)=0$.

The most remarkable feature of the vector-like dark energy is that its EoS
$w_A\equiv p_A/\rho_A$ can be smaller than $-1$, while possessing a conventional
positive kinetic term. This is thanks to the additional term in proportion to $dV/dA^2$
in $p_A$. While the energy density $\rho_A$ is positive, one can find that the condition
for $w_A<-1$ is
\be{eq11}
\frac{dV}{dA^2}A^2<-\left(\dot{A}+HA\right)^2.
\ee
Thus, $dV/dA^2<0$ is necessary~\cite{r46,r49}. There are other interesting issues concerning
the vector-like dark energy. We refer to the original paper~\cite{r46} for more details.


\section{\label{sec3} Dynamical system of vector-like dark energy interacting with
background perfect fluid}
As mentioned in Sec.~\ref{sec1}, in this work we will generalize the original vector-like dark energy
model~\cite{r46} to include the interaction between the vector-like dark energy and background
matter. The background matter is described by a perfect fluid with  barotropic equation of state
\be{eq12}
p_m=w_m\rho_m\equiv (\gamma-1)\rho_m,
\ee
where the barotropic index $\gamma$ is a constant and satisfies $0<\gamma\leq 2$. In particular,
$\gamma=1$ and $4/3$ correspond to dust matter and radiation, respectively. We assume the
vector-like dark energy and background matter interact through an interaction term $C$, according to
\bea
&&\dot{\rho}_A+3H\left(\rho_A+p_A\right)=-C,\label{eq13}\\
&&\dot{\rho}_m+3H\left(\rho_m+p_m\right)=C,\label{eq14}
\eea
which preserves the total energy conservation equation
$\dot{\rho}_{tot}+3H\left(\rho_{tot}+p_{tot}\right)=0$. It is worth noting that the equation of
motion~(\ref{eq6}) should be changed when $C\not=0$, a new term due to $C$ will appear in its
right hand side.

Following~\cite{r50,r51,r52}, we introduce following dimensionless variables
\be{eq15}
x\equiv\frac{\kappa\dot{A}}{\sqrt{6}H},~~~~y\equiv\frac{\kappa\sqrt{V}}{\sqrt{3}H},~~~~
z\equiv\frac{\kappa\sqrt{\rho_m}}{\sqrt{3}H},~~~~u\equiv\frac{\kappa A}{\sqrt{6}}.
\ee
By the help of Eqs.~(\ref{eq7})--(\ref{eq10}), the evolution equations~(\ref{eq13}) and (\ref{eq14}) can
then be rewritten as a dynamical system~\cite{r48}, i.e.
\bea
&&x^\prime=6\left[\left(x+u\right)^2+\frac{\gamma}{4}z^2+\Theta\right]\left(x+u\right)-2\Theta u^{-1}
-3x-2u-C_1,\label{eq16}\\
&&y^\prime=6y\left[\left(x+u\right)^2+\frac{\gamma}{4}z^2+\left(1+
\frac{1}{3}xy^{-2}u^{-1}\right)\Theta\right],\label{eq17}\\
&&z^\prime=6z\left[\left(x+u\right)^2+\frac{\gamma}{4}z^2+\Theta-\frac{\gamma}{4}\right]+C_2,\label{eq18}\\
&&u^\prime=x,\label{eq19}
\eea
where
\be{eq20}
\Theta\equiv\frac{u^2}{H^2}\frac{dV}{dA^2},
\ee
and
\be{eq21}
C_1\equiv\frac{\kappa^2 C}{18H^3}\left(x+u\right)^{-1},~~~~~~~
C_2\equiv\frac{z\, C}{2H\rho_m},
\ee
a prime denotes derivative with respect to the so-called $e$-folding time $N\equiv\ln a$,
and we have used
\be{eq22}
-\frac{\dot{H}}{H^2}=6\left[\left(x+u\right)^2+\frac{\gamma}{4}z^2+\Theta\right].
\ee
The Friedmann constraint Eq.~(\ref{eq7}) becomes
\be{eq23}
3\left[\left(x+u\right)^2+y^2\right]+z^2=1.
\ee
The fractional energy densities of vector-like dark energy and background matter
are given by
\be{eq24}
\Omega_A\equiv\frac{\kappa^2\rho_A}{3H^2}=3\left[\left(x+u\right)^2+y^2\right],~~~~~~~
\Omega_m\equiv\frac{\kappa^2\rho_m}{3H^2}=z^2,
\ee
respectively. The EoS of vector-like dark energy and the effective EoS of the whole system are
\be{eq25}
w_A\equiv\frac{p_A}{\rho_A}=\frac{\left(x+u\right)^2-3y^2+4\Theta}{3\left[\left(x+u\right)^2+y^2\right]},
\ee
and
\be{eq26}
w_{eff}\equiv\frac{p_{tot}}{\rho_{tot}}=\Omega_A w_A+\Omega_m w_m,
\ee
respectively. From Eq.~(\ref{eq25}), it is easy to see that the condition for $w_A<-1$ is
\be{eq27}
\left(x+u\right)^2+\Theta<0,
\ee
which is equivalent to Eq.~(\ref{eq11}). Finally, it is worth noting that $y\geq 0$ and
$z\geq 0$ by definition, and in what follows, we only consider the case of expanding universe
with $H>0$.

It is easy to see that Eqs.~(\ref{eq16})--(\ref{eq19}) become an autonomous system when the
potential $V\left(A^2\right)$ is chosen to be an inverse power-law or exponential potential and
the interaction term $C$ is chosen to be a suitable form. Indeed, we will consider the model
with an inverse power-law and exponential potential in Sec.~\ref{sec4} and \ref{sec5}, respectively.
In each model with different potential, we consider four cases with different interaction forms
between vector-like dark energy and background perfect fluid, which are taken as the most
familiar interaction terms extensively considered in the literature~\cite{r51,r52,r53,r54,r55,r56}, namely
\begin{eqnarray*}
&{\rm Case~(I)} &C=3\alpha H\rho_m,\\
&{\rm Case~(II)} &C=3\beta H\rho_{tot}=3\beta H\left(\rho_A+\rho_m\right),\\
&{\rm Case~(III)} &C=\eta\kappa\rho_m\dot{A},\\
&{\rm Case~(IV)} &C=3\sigma H\rho_A,
\end{eqnarray*}
where $\alpha$, $\beta$, $\eta$ and $\sigma$ are dimensionless constants.

In the next two sections, we firstly obtain  the critical points $(\bar{x},\bar{y},\bar{z},\bar{u})$ of
the autonomous system by imposing the conditions
$\bar{x}^\prime=\bar{y}^\prime=\bar{z}^\prime=\bar{u}^\prime=0$. Of course, they are subject to
the Friedmann constraint, namely $3\left[\left(\bar{x}+\bar{u}\right)^2+\bar{y}^2\right]+\bar{z}^2=1$.
We then discuss the existence and stability of these critical points. An attractor is one of the stable
critical points of the autonomous system.


\section{\label{sec4} Model with inverse power-law potential}
In this section, we consider the vector-like dark energy model with an inverse power-law potential
\be{eq28}
V\left(A^2\right)=V_0\left(\kappa^2A^2\right)^{-n},
\ee
where $n$ is a positive dimensionless constant (required by the condition $dV/dA^2<0$). In this case,
\be{eq29}
\Theta=-\frac{n}{2}y^2.
\ee
One can obtain the critical points $(\bar{x},\bar{y},\bar{z},\bar{u})$ of the dynamical system
Eqs.~(\ref{eq16})--(\ref{eq19}) with Eqs.~(\ref{eq28}) and (\ref{eq29}) by imposing the conditions
$\bar{x}^\prime=\bar{y}^\prime=\bar{z}^\prime=\bar{u}^\prime=0$. Note that these critical points must
satisfy the Friedmann constraint~(\ref{eq23}), $\bar{y}\geq 0$, $\bar{z}\geq 0$ and the requirement
of $\bar{x},\,\bar{y},\,\bar{z},\,\bar{u}$ all being real. To study the stability of these
critical points, we substitute linear perturbations $x\to\bar{x}+\delta x$, $y\to\bar{y}+\delta y$,
$z\to\bar{z}+\delta z$, and $u\to\bar{u}+\delta u$ about the critical point
$(\bar{x},\bar{y},\bar{z},\bar{u})$ into dynamical system Eqs.~(\ref{eq16})--(\ref{eq19}) with
Eqs.~(\ref{eq28}) and (\ref{eq29}) and linearize them. Because of the Friedmann constraint~(\ref{eq23}),
there are only three independent evolution equations, i.e.
\bea
&\delta x^\prime=&\left[9(n+2)(\bar{x}+\bar{u})^2-2n\bar{u}^{-1}(\bar{x}+\bar{u})+
\left(\frac{3}{2}\gamma+n\right)\bar{z}^2-n-3\right]\delta x\nonumber\\
& &+2\bar{z}\left[\left(\frac{3}{2}\gamma+n\right)(\bar{x}+\bar{u})-\frac{n}{3}\bar{u}^{-1}\right]\delta z
+\left[\left(18+9n+n\bar{u}^{-2}\right)\left(\bar{x}+\bar{u}\right)^2+\left(\frac{3}{2}\gamma
+n\right)\bar{z}^2\right.\nonumber\\
& &\left.-\frac{n}{3}\left(1-\bar{z}^2\right)\bar{u}^{-2}-2n\bar{x}\bar{u}^{-1}-3n-2\right]\delta u
-\delta C_1,\label{eq30}\\
&\delta z^\prime=&6(n+2)\bar{z}(\bar{x}+\bar{u})\delta x+\left[3(n+2)(\bar{x}+\bar{u})^2
+\left(\frac{3}{2}\gamma+n\right)\left(3\bar{z}^2-1\right)\right]\delta z\nonumber\\
& &+6(n+2)\bar{z}(\bar{x}+\bar{u})\delta u+\delta C_2,\label{eq31}\\
&\delta u^\prime=&\delta x,\label{eq32}
\eea
where $\delta C_1$ and $\delta C_2$ are the linear perturbations coming from $C_1$ and $C_2$,
respectively. The three eigenvalues of the coefficient matrix of the above equations determine the
stability of the corresponding critical point.


\subsection{Case~(I)~$C=3\alpha H\rho_m$}

In this case, $C_1=\frac{\alpha}{2}(x+u)^{-1}z^2$ and $C_2=\frac{3}{2}\alpha z$. The physically
reasonable critical points of the dynamical system Eqs.~(\ref{eq16})--(\ref{eq19}) with
Eqs.~(\ref{eq28}) and (\ref{eq29}) are summarized in Table~\ref{tab1}. Next, we consider the
stability of these critical points. Substituting $\delta C_1=-\frac{\alpha}{2}\bar{z}^2
(\bar{x}+\bar{u})^{-2}\delta x+\alpha\bar{z}(\bar{x}+\bar{u})^{-1}\delta z-\frac{\alpha}{2}\bar{z}^2
(\bar{x}+\bar{u})^{-2}\delta u$, $\delta C_2=\frac{3}{2}\alpha\delta z$, and the corresponding critical
point $(\bar{x},\bar{y},\bar{z},\bar{u})$ into Eqs.~(\ref{eq30})--(\ref{eq32}), we find that Point~(P.I.1)
is always unstable; Point~(P.I.2) exists and is stable under condition $\alpha<\gamma$; Point~(P.I.3)
is unstable if it can exist. The unique late time attractor~(P.I.2) has
\be{eq33}
\Omega_A=1,~~~~\Omega_m=0,~~~~w_A=-1,~~~~w_{eff}=-1,
\ee
which is a vector-like dark energy dominated solution.

\begin{table}[htbp]
\begin{center}
\begin{tabular}{c|c}
\hline\hline
Label & Critical Point $(\bar{x},\bar{y},\bar{z},\bar{u})$ \\ \hline
P.I.1 & \ 0,\ 0,\ 0,\ $\pm\frac{1}{\sqrt{3}}$ \ \\
P.I.2 & \ 0,\ $\sqrt{\frac{2}{3(2+n)}}$,\ 0,\  $\pm\sqrt{\frac{n}{3(2+n)}}$ \ \\
P.I.3 & \ 0,\ 0,\ $\sqrt{1-\frac{3\alpha}{3\gamma-4}}$,\ $\pm\sqrt{\frac{\alpha}{3\gamma-4}}$ \ \\ \hline\hline
\end{tabular}
\end{center}
\caption{\label{tab1} Critical points for Case~(I) $C=3\alpha H\rho_m$ in the model with
inverse power-law potential.}
\end{table}


\subsection{Case~(II)~$C=3\beta H\rho_{tot}=3\beta H\left(\rho_A+\rho_m\right)$}
In this case, $C_1=\frac{\beta}{2}(x+u)^{-1}$ and
$C_2=\frac{3}{2}\beta z^{-1}$. We present the physically
reasonable critical points of the dynamical system
Eqs.~(\ref{eq16})--(\ref{eq19}) with Eqs.~(\ref{eq28}) and
(\ref{eq29}) in Table~\ref{tab2}, where \be{eq34}
r_1\equiv\sqrt{1+\frac{12\beta}{4-3\gamma}}. \ee Then, we
substitute $\delta
C_1=-\frac{\beta}{2}(\bar{x}+\bar{u})^{-2}\left(\delta x+\delta
u\right)$, $\delta C_2=-\frac{3}{2}\beta\bar{z}^{-2}\delta z$, and
the corresponding critical point
$(\bar{x},\bar{y},\bar{z},\bar{u})$ into
Eqs.~(\ref{eq30})--(\ref{eq32}) to study its stability. We find
that Point~(P.II.1) exists and is stable under conditions
$0<\beta<\frac{4-3\gamma}{12}\left\{\left[\frac{4+3\gamma}{3(3\gamma-4)}\right]^2-1\right\}$
and $\gamma>8/3$, which is out of the range $0<\gamma\leq 2$;
Point~(P.II.2) exists and is stable under conditions
$\beta<\min\left\{0,\,\frac{4-3\gamma}{12}\left[\frac{(4+3\gamma)^2}{9(4-3\gamma)^2}-1\right]\right\}$
and $\gamma<4/3$; Point~(P.II.3) exists and is stable in a proper
parameter-space~\cite{r57}.

The late time attractor~(P.II.2) has
\be{eq35}
\Omega_A=\frac{1}{2}(1-r_1),~~~~\Omega_m=\frac{1}{2}(1+r_1),~~~~
w_A=\frac{1}{3},~~~~w_{eff}=\frac{1}{6}\left[\left(1+r_1\right)\left(3\gamma-4\right)+2\right],
\ee which is a scaling solution. The late time attractor~(P.II.3)
has
\be{eq36}
\Omega_A=1-\frac{\beta}{\gamma},~~~~\Omega_m=\frac{\beta}{\gamma},~~~~
w_A=-1-\frac{\beta\gamma}{\gamma-\beta},~~~~w_{eff}=-1,
\ee
which is a scaling solution also. Note that $w_{eff}$ of
attractor~(P.II.2) is larger than $-1$, since $0<\gamma\leq 2$ and
$0<r_1<1$ which is required by Eq.~(\ref{eq34}) and its
corresponding $\Omega_m$, while $w_A$ of attractor~(P.II.3) is
smaller than $-1$, since $0<\beta<\gamma$ is required by its
corresponding $\Omega_m$.

\begin{table}[htbp]
\begin{center}
\begin{tabular}{c|c}
\hline\hline
Label & Critical Point $(\bar{x},\bar{y},\bar{z},\bar{u})$ \\ \hline
P.II.1 & \ 0,\ 0,\ $\left[\frac{1}{2}(1-r_1)\right]^{1/2}$,\ $\pm\left[\frac{1}{6}(1+r_1)\right]^{1/2}$ \ \\
P.II.2 & \ 0,\ 0,\ $\left[\frac{1}{2}(1+r_1)\right]^{1/2}$,\ $\pm\left[\frac{1}{6}(1-r_1)\right]^{1/2}$ \ \\
P.II.3 & \ 0,\ $\sqrt{\frac{4\gamma+\beta(-4+3\gamma)}{6(2+n)\gamma}}$,\ $\sqrt{\frac{\beta}{\gamma}}$,\
$\pm\sqrt{\frac{2n(\gamma-\beta)-3\beta\gamma}{6(2+n)\gamma}}$ \ \\ \hline\hline
\end{tabular}
\end{center}
\caption{\label{tab2} Critical points for Case~(II) $C=3\beta H\rho_{tot}=3\beta H\left(\rho_A+\rho_m\right)$
in the model with inverse power-law potential. $r_1$ is given in Eq.~(\ref{eq34}).}
\end{table}


\subsection{Case~(III)~$C=\eta\kappa\rho_m\dot{A}$}
In this case, $C_1=\frac{\eta}{\sqrt{6}}(x+u)^{-1}xz^2$ and $C_2=\sqrt{\frac{3}{2}}\eta xz$. The physically
reasonable critical points of the dynamical system Eqs.~(\ref{eq16})--(\ref{eq19}) with Eqs.~(\ref{eq28})
and (\ref{eq29}) are shown in Table~\ref{tab3}. Substituting $\delta C_1=\frac{\eta}{\sqrt{6}}(\bar{x}+
\bar{u})^{-1}\bar{z}^2\left[1-(\bar{x}+\bar{u})^{-1}\bar{x}\right]\delta x+\sqrt{\frac{2}{3}}\eta(\bar{x}+
\bar{u})^{-1}\bar{x}\bar{z}\delta z-\frac{\eta}{\sqrt{6}}(\bar{x}+\bar{u})^{-2}\bar{x}\bar{z}^2\delta u$,
$\delta C_2=\sqrt{\frac{3}{2}}\eta\bar{z}\delta x+\sqrt{\frac{3}{2}}\eta\bar{x}\delta z$, and the corresponding
critical point $(\bar{x},\bar{y},\bar{z},\bar{u})$ into Eqs.~(\ref{eq30})--(\ref{eq32}), we find that
Point~(P.III.1) is always unstable, while Point~(P.III.2) is always stable. The unique late time
attractor~(P.III.2) has
\be{eq37}
\Omega_A=1,~~~~\Omega_m=0,~~~~w_A=-1,~~~~w_{eff}=-1,
\ee
which is a vector-like dark energy dominated solution.

\begin{table}[htbp]
\begin{center}
\begin{tabular}{c|c}
\hline\hline
Label & Critical Point $(\bar{x},\bar{y},\bar{z},\bar{u})$ \\ \hline
P.III.1 & \ 0,\ 0,\ 0,\ $\pm\frac{1}{\sqrt{3}}$ \ \\
P.III.2 & \ 0,\ $\sqrt{\frac{2}{3(2+n)}}$,\ 0,\ $\pm\sqrt{\frac{n}{3(2+n)}}$ \ \\ \hline\hline
\end{tabular}
\end{center}
\caption{\label{tab3} Critical points for Case~(III) $C=\eta\kappa\rho_m\dot{A}$
in the model with inverse power-law potential.}
\end{table}


\subsection{Case~(IV)~$C=3\sigma H\rho_A$}
In this case,
$C_1=\frac{3}{2}\sigma\left[(x+u)^2+y^2\right](x+u)^{-1}=\frac{\sigma}{2}(1-z^2)(x+u)^{-1}$
and
$C_2=\frac{9}{2}\sigma\left[(x+u)^2+y^2\right]z^{-1}=\frac{3}{2}\sigma\left(z^{-1}-z\right)$.
The physically reasonable critical points of the dynamical system
Eqs.~(\ref{eq16})--(\ref{eq19}) with Eqs.~(\ref{eq28}) and
(\ref{eq29}) are shown in Table~\ref{tab4}, where \be{eq38}
r_2\equiv\frac{\sigma}{-4+3\gamma}. \ee Next, we consider the
stability of these critical points. Substituting $\delta
C_1=-\frac{\sigma}{2}
(\bar{x}+\bar{u})^{-2}\left(1-\bar{z}^2\right)\delta
x-\sigma\bar{z}(\bar{x}+\bar{u})^{-1}\delta z-\frac{\sigma}{2}
(\bar{x}+\bar{u})^{-2}\left(1-\bar{z}^2\right)\delta u$, $\delta
C_2=-\frac{3}{2}\sigma\left(\bar{z}^{-2}+1\right)\delta z$, and
the corresponding critical point
$(\bar{x},\bar{y},\bar{z},\bar{u})$ into
Eqs.~(\ref{eq30})--(\ref{eq32}), we find that Point~(P.IV.1) is
unstable if it can exist, while Point~(P.IV.2) exists and is
stable in a proper parameter-space~\cite{r57}. The unique late
time attractor~(P.IV.2) has \be{eq39}
\Omega_A=\frac{\gamma}{\gamma+\sigma},~~~~\Omega_m=\frac{\sigma}{\gamma+\sigma},~~~~
w_A=-1-\sigma,~~~~w_{eff}=-1, \ee which is a scaling solution.
Note that $w_A$ of attractor~(P.IV.2) is smaller than $-1$, since
$\sigma>0$ is required by its corresponding $\Omega_m$.

\begin{table}[htbp]
\begin{center}
\begin{tabular}{c|c}
\hline\hline
Label & Critical Point $(\bar{x},\bar{y},\bar{z},\bar{u})$ \\ \hline
P.IV.1 & \ 0,\ 0,\ $(3r_2)^{1/2}$,\ $\pm\left(\frac{1}{3}-r_2\right)^{1/2}$ \ \\
P.IV.2 & \ 0,\ $\sqrt{\frac{\gamma(4+3\sigma)}{6(2+n)(\gamma+\sigma)}}$,\
$\sqrt{\frac{\sigma}{\gamma+\sigma}}$,\
$\pm\sqrt{\frac{\gamma(2n-3\sigma)}{6(2+n)(\gamma+\sigma)}}$ \ \\ \hline\hline
\end{tabular}
\end{center}
\caption{\label{tab4} Critical points for Case~(IV) $C=3\sigma H\rho_A$
in the model with inverse power-law potential. $r_2$ is given in Eq.~(\ref{eq38}).}
\end{table}


\section{\label{sec5} Model with exponential potential}
In this section, we consider the vector-like dark energy model with an exponential potential
\be{eq40}
V\left(A^2\right)=V_0\exp\left(-\lambda\kappa^2A^2\right),
\ee
where $\lambda$ is a positive dimensionless constant (required by the condition $dV/dA^2<0$).
In this case,
\be{eq41}
\Theta=-3\lambda u^2 y^2.
\ee
One can obtain the critical points $(\bar{x},\bar{y},\bar{z},\bar{u})$ of the dynamical system
Eqs.~(\ref{eq16})--(\ref{eq19}) with Eqs.~(\ref{eq40}) and (\ref{eq41}) by imposing the conditions
$\bar{x}^\prime=\bar{y}^\prime=\bar{z}^\prime=\bar{u}^\prime=0$. Note that these critical points must
satisfy the Friedmann constraint~(\ref{eq23}), $\bar{y}\geq 0$, $\bar{z}\geq 0$ and the requirement
of $\bar{x},\,\bar{y},\,\bar{z},\,\bar{u}$ all being real. To study the stability of these
critical points, we substitute linear perturbations $x\to\bar{x}+\delta x$, $y\to\bar{y}+\delta y$,
$z\to\bar{z}+\delta z$, and $u\to\bar{u}+\delta u$ about the critical point
$(\bar{x},\bar{y},\bar{z},\bar{u})$ into dynamical system Eqs.~(\ref{eq16})--(\ref{eq19}) with
Eqs.~(\ref{eq40}) and (\ref{eq41}) and linearize them. Because of the Friedmann constraint~(\ref{eq23}),
there are only three independent evolution equations, namely
\bea
&\delta x^\prime=&6\left[3\left(1+3\lambda\bar{u}^2\right)(\bar{x}+\bar{u})^2
-2\lambda\bar{u}(\bar{x}+\bar{u})+\left(\lambda\bar{u}^2+\frac{\gamma}{4}\right)\bar{z}^2
-\lambda\bar{u}^2-\frac{1}{2}\right]\delta x\nonumber\\
& &\left.+4\bar{z}\left[3(\bar{x}+\bar{u})\left(\lambda\bar{u}^2+\frac{\gamma}{4}\right)
-\lambda\bar{u}\right]\delta z+2\right\{18\lambda\bar{u}(\bar{x}+\bar{u})^3
+3\left(9\lambda\bar{u}^2+3-\lambda\right)(\bar{x}+\bar{u})^2\nonumber\\
& &\left.-6\lambda\bar{u}(\bar{x}+\bar{u})+\left[6\lambda\bar{u}(\bar{x}+\bar{u})+
3\left(\lambda\bar{u}^2+\frac{\gamma}{4}\right)-\lambda\right]\left(\bar{z}^2-1\right)
+\frac{3}{4}\gamma-1\right\}\delta u-\delta C_1,\label{eq42}\\
&\delta z^\prime=&12\bar{z}\left(1+3\lambda\bar{u}^2\right)(\bar{x}+\bar{u})\delta x
+6\left[\left(1+3\lambda\bar{u}^2\right)(\bar{x}+\bar{u})^2+\left(\lambda\bar{u}^2
+\frac{\gamma}{4}\right)\left(3\bar{z}^2-1\right)\right]\delta z\nonumber\\
& &+12\bar{z}\left[3\lambda\bar{u}(\bar{x}+\bar{u})^2+\left(1+3\lambda\bar{u}^2\right)(\bar{x}+\bar{u})
+\lambda\bar{u}\left(\bar{z}^2-1\right)\right]\delta u+\delta C_2,\label{eq43}\\
&\delta u^\prime=&\delta x,\label{eq44}
\eea
where $\delta C_1$ and $\delta C_2$ are the linear perturbations coming from $C_1$ and $C_2$,
respectively. The three eigenvalues of the coefficient matrix of the above equations determine the
stability of the corresponding critical point.


\subsection{Case~(I)~$C=3\alpha H\rho_m$}
In this case, $C_1=\frac{\alpha}{2}(x+u)^{-1}z^2$ and $C_2=\frac{3}{2}\alpha z$. The physically
reasonable critical points of the dynamical system Eqs.~(\ref{eq16})--(\ref{eq19}) with
Eqs.~(\ref{eq40}) and (\ref{eq41}) are summarized in Table~\ref{tab5}. Next, we consider the
stability of these critical points. Substituting $\delta C_1=-\frac{\alpha}{2}\bar{z}^2
(\bar{x}+\bar{u})^{-2}\delta x+\alpha\bar{z}(\bar{x}+\bar{u})^{-1}\delta z-\frac{\alpha}{2}\bar{z}^2
(\bar{x}+\bar{u})^{-2}\delta u$, $\delta C_2=\frac{3}{2}\alpha\delta z$, and the corresponding critical
point $(\bar{x},\bar{y},\bar{z},\bar{u})$ into Eqs.~(\ref{eq42})--(\ref{eq44}), we find that Point~(E.I.1)
is always unstable; Point~(E.I.2) exists and is stable under conditions $\alpha<\gamma$ and $\lambda>1$;
Point~(E.I.3) is unstable if it can exist. The unique late time attractor~(E.I.2) has
\be{eq45}
\Omega_A=1,~~~~\Omega_m=0,~~~~w_A=-1,~~~~w_{eff}=-1,
\ee
which is a vector-like dark energy dominated solution.

\begin{table}[htbp]
\begin{center}
\begin{tabular}{c|c}
\hline\hline
Label & Critical Point $(\bar{x},\bar{y},\bar{z},\bar{u})$ \\ \hline
E.I.1 & \ 0,\ 0,\ 0,\ $\pm\frac{1}{\sqrt{3}}$ \ \\
E.I.2 & \ 0,\ $\frac{1}{\sqrt{3\lambda}}$,\ 0,\  $\pm\sqrt{\frac{\lambda-1}{3\lambda}}$ \ \\
E.I.3 & \ 0,\ 0,\ $\sqrt{1-\frac{3\alpha}{3\gamma-4}}$,\ $\pm\sqrt{\frac{\alpha}{3\gamma-4}}$ \ \\ \hline\hline
\end{tabular}
\end{center}
\caption{\label{tab5} Critical points for Case~(I) $C=3\alpha H\rho_m$ in the model with
exponential potential.}
\end{table}


\subsection{Case~(II)~$C=3\beta H\rho_{tot}=3\beta H\left(\rho_A+\rho_m\right)$}
In this case, $C_1=\frac{\beta}{2}(x+u)^{-1}$ and
$C_2=\frac{3}{2}\beta z^{-1}$. We present the physically
reasonable critical points of the dynamical system
Eqs.~(\ref{eq16})--(\ref{eq19}) with Eqs.~(\ref{eq40}) and
(\ref{eq41}) in Table~\ref{tab6}, where \be{eq46}
r_3\equiv\sqrt{-3\beta\gamma^2\lambda+(\gamma+\beta\lambda-\gamma\lambda)^2}.
\ee Then, we substitute $\delta
C_1=-\frac{\beta}{2}(\bar{x}+\bar{u})^{-2}\left(\delta x+\delta
u\right)$, $\delta C_2=-\frac{3}{2}\beta\bar{z}^{-2}\delta z$, and
the corresponding critical point
$(\bar{x},\bar{y},\bar{z},\bar{u})$ into
Eqs.~(\ref{eq42})--(\ref{eq44}) to study its stability. We find
that Point~(E.II.1) exists and is stable under conditions
$0<\beta<\frac{4-3\gamma}{12}\left\{\left[\frac{4+3\gamma}{3(3\gamma-4)}\right]^2-1\right\}$
and $\gamma>8/3$, which is out of the range $0<\gamma\leq 2$;
Point~(E.II.2) exists and is stable under conditions
$\beta<\min\left\{0,\,\frac{4-3\gamma}{12}\left[\frac{(4+3\gamma)^2}{9(4-3\gamma)^2}-1\right]\right\}$
and $\gamma<4/3$; Points~(E.II.3) and (E.II.4) exist and are
stable in a proper parameter-space~\cite{r57}, respectively.

The late time attractor~(E.II.2) has
\be{eq47}
\Omega_A=\frac{1}{2}(1-r_1),~~~~\Omega_m=\frac{1}{2}(1+r_1),~~~~
w_A=\frac{1}{3},~~~~w_{eff}=\frac{1}{6}\left[\left(1+r_1\right)\left(3\gamma-4\right)+2\right],
\ee
which is a scaling solution. The late time attractors~(E.II.3) and (E.II.4) both have
\be{eq48}
\Omega_A=1-\frac{\beta}{\gamma},~~~~\Omega_m=\frac{\beta}{\gamma},~~~~
w_A=-1-\frac{\beta\gamma}{\gamma-\beta},~~~~w_{eff}=-1,
\ee
which are scaling solutions also. Note that $w_{eff}$ of attractor~(E.II.2) is larger than $-1$, since
$0<\gamma\leq 2$ and $0<r_1<1$ which is required by Eq.~(\ref{eq34}) and its corresponding $\Omega_m$,
while $w_A$ of attractors~(E.II.3) and (E.II.4) are both smaller than $-1$, since $0<\beta<\gamma$ is
required by their corresponding $\Omega_m$.

\begin{table}[htbp]
\begin{center}
\begin{tabular}{c|c}
\hline\hline
Label & Critical Point $(\bar{x},\bar{y},\bar{z},\bar{u})$ \\ \hline
E.II.1 & \ 0,\ 0,\ $\left[\frac{1}{2}(1-r_1)\right]^{1/2}$,\ $\pm\left[\frac{1}{6}(1+r_1)\right]^{1/2}$ \ \\
E.II.2 & \ 0,\ 0,\ $\left[\frac{1}{2}(1+r_1)\right]^{1/2}$,\ $\pm\left[\frac{1}{6}(1-r_1)\right]^{1/2}$ \ \\
E.II.3 & \ 0,\ $\sqrt{\frac{\gamma-\beta\lambda+\gamma\lambda-r_3}{6\gamma\lambda}}$,\
$\sqrt{\frac{\beta}{\gamma}}$,\
$\pm\sqrt{\frac{\gamma(-1+\lambda)-\beta\lambda+r_3}{6\gamma\lambda}}$ \ \\
E.II.4 & \ 0,\ $\sqrt{\frac{\gamma-\beta\lambda+\gamma\lambda+r_3}{6\gamma\lambda}}$,\
$\sqrt{\frac{\beta}{\gamma}}$,\
$\pm\sqrt{\frac{\gamma(-1+\lambda)-\beta\lambda-r_3}{6\gamma\lambda}}$ \ \\ \hline\hline
\end{tabular}
\end{center}
\caption{\label{tab6} Critical points for Case~(II) $C=3\beta H\rho_{tot}=3\beta H\left(\rho_A+\rho_m\right)$
in the model with exponential potential. $r_1$ and $r_3$ are given in Eqs.~(\ref{eq34})
and (\ref{eq46}), respectively.}
\end{table}


\subsection{Case~(III)~$C=\eta\kappa\rho_m\dot{A}$}
In this case, $C_1=\frac{\eta}{\sqrt{6}}(x+u)^{-1}xz^2$ and $C_2=\sqrt{\frac{3}{2}}\eta xz$. The physically
reasonable critical points of the dynamical system Eqs.~(\ref{eq16})--(\ref{eq19}) with Eqs.~(\ref{eq40})
and (\ref{eq41}) are shown in Table~\ref{tab7}. Substituting $\delta C_1=\frac{\eta}{\sqrt{6}}(\bar{x}+
\bar{u})^{-1}\bar{z}^2\left[1-(\bar{x}+\bar{u})^{-1}\bar{x}\right]\delta x+\sqrt{\frac{2}{3}}\eta(\bar{x}+
\bar{u})^{-1}\bar{x}\bar{z}\delta z-\frac{\eta}{\sqrt{6}}(\bar{x}+\bar{u})^{-2}\bar{x}\bar{z}^2\delta u$,
$\delta C_2=\sqrt{\frac{3}{2}}\eta\bar{z}\delta x+\sqrt{\frac{3}{2}}\eta\bar{x}\delta z$, and the corresponding
critical point $(\bar{x},\bar{y},\bar{z},\bar{u})$ into Eqs.~(\ref{eq42})--(\ref{eq44}), we find that
Point~(E.III.1) is always unstable, while Point~(E.III.2) exists and is stable under condition $\lambda>1$.
The unique late time attractor~(E.III.2) has
\be{eq49}
\Omega_A=1,~~~~\Omega_m=0,~~~~w_A=-1,~~~~w_{eff}=-1,
\ee
which is a vector-like dark energy dominated solution.

\begin{table}[htbp]
\begin{center}
\begin{tabular}{c|c}
\hline\hline
Label & Critical Point $(\bar{x},\bar{y},\bar{z},\bar{u})$ \\ \hline
E.III.1 & \ 0,\ 0,\ 0,\ $\pm\frac{1}{\sqrt{3}}$ \ \\
E.III.2 & \ 0,\ $\frac{1}{\sqrt{3\lambda}}$,\ 0,\ $\pm\sqrt{\frac{\lambda-1}{3\lambda}}$ \ \\ \hline\hline
\end{tabular}
\end{center}
\caption{\label{tab7} Critical points for Case~(III) $C=\eta\kappa\rho_m\dot{A}$
in the model with exponential potential.}
\end{table}


\subsection{Case~(IV)~$C=3\sigma H\rho_A$}
In this case,
$C_1=\frac{3}{2}\sigma\left[(x+u)^2+y^2\right](x+u)^{-1}=\frac{\sigma}{2}(1-z^2)(x+u)^{-1}$
and
$C_2=\frac{9}{2}\sigma\left[(x+u)^2+y^2\right]z^{-1}=\frac{3}{2}\sigma\left(z^{-1}-z\right)$.
The physically reasonable critical points of the dynamical system
Eqs.~(\ref{eq16})--(\ref{eq19}) with Eqs.~(\ref{eq40}) and
(\ref{eq41}) are shown in Table~\ref{tab8}, where \be{eq50}
r_4\equiv\sqrt{-3\gamma\lambda\sigma(\gamma+\sigma)+(\gamma-\gamma\lambda+\sigma)^2}.
\ee Then, we consider the stability of these critical points.
Substituting $\delta C_1=-\frac{\sigma}{2}
(\bar{x}+\bar{u})^{-2}\left(1-\bar{z}^2\right)\delta
x-\sigma\bar{z}(\bar{x}+\bar{u})^{-1}\delta z-\frac{\sigma}{2}
(\bar{x}+\bar{u})^{-2}\left(1-\bar{z}^2\right)\delta u$, $\delta
C_2=-\frac{3}{2}\sigma\left(\bar{z}^{-2}+1\right)\delta z$, and
the corresponding critical point
$(\bar{x},\bar{y},\bar{z},\bar{u})$ into
Eqs.~(\ref{eq42})--(\ref{eq44}), we find that Point~(E.IV.1) is
unstable if it can exist, while Points~(E.IV.2) and (E.IV.3) exist
and are stable in a proper parameter-space~\cite{r57},
respectively. The late time attractors~(E.IV.2) and (E.IV.3) both
have \be{eq51}
\Omega_A=\frac{\gamma}{\gamma+\sigma},~~~~\Omega_m=\frac{\sigma}{\gamma+\sigma},~~~~
w_A=-1-\sigma,~~~~w_{eff}=-1, \ee which are scaling solutions.
Note that $w_A$ of attractors~(E.IV.2) and (E.IV.3) are both
smaller than $-1$, since $\sigma>0$ is required by their
corresponding $\Omega_m$.

\begin{table}[htbp]
\begin{center}
\begin{tabular}{c|c}
\hline\hline
Label & Critical Point $(\bar{x},\bar{y},\bar{z},\bar{u})$ \\ \hline
E.IV.1 & \ 0,\ 0,\ $(3r_2)^{1/2}$,\ $\pm\left(\frac{1}{3}-r_2\right)^{1/2}$ \ \\
E.IV.2 & \ 0,\ $\sqrt{\frac{\gamma(1+\lambda)+\sigma-r_4}{6\lambda(\gamma+\sigma)}}$,\
$\sqrt{\frac{\sigma}{\gamma+\sigma}}$,\
$\pm\sqrt{\frac{\gamma(-1+\lambda)-\sigma+r_4}{6\lambda(\gamma+\sigma)}}$ \ \\
E.IV.3 & \ 0,\ $\sqrt{\frac{\gamma(1+\lambda)+\sigma+r_4}{6\lambda(\gamma+\sigma)}}$,\
$\sqrt{\frac{\sigma}{\gamma+\sigma}}$,\
$\pm\sqrt{\frac{\gamma(-1+\lambda)-\sigma-r_4}{6\lambda(\gamma+\sigma)}}$ \ \\ \hline\hline
\end{tabular}
\end{center}
\caption{\label{tab8} Critical points for Case~(IV) $C=3\sigma H\rho_A$
in the model with exponential potential. $r_2$ and $r_4$ are given in Eqs.~(\ref{eq38})
and (\ref{eq50}), respectively.}
\end{table}


\section{\label{sec6} The first and second cosmological coincidence problems}
As is well known, the most frequently used approach to alleviate the (first) cosmological coincidence
problem is the scaling attractor(s) in the interacting dark energy scenario (see~\cite{r48,r50,r51,r52,r53,
r54,r55,r56} for examples). The dark energy can exchange energy with the background matter (usually
the cold dark matter), through the interaction between them. The most desirable feature of dynamical
system is that the whole system will eventually evolve to its attractors, having nothing to do with the
initial conditions. Therefore, fine-tunings are needless. When the universe is attracted into the scaling
attractor, a balance can be achieved, thanks to the interaction. In the scaling attractor, the effective
densities of dark energy and background matter decrease in the same manner with the expansion of
our universe, and the ratio of dark energy and background matter becomes a constant. So, it is not
strange that we are living in an epoch when the densities of dark energy and matter are comparable.
In this sense, the (first) cosmological coincidence problem is alleviated (see~\cite{r48,r50,r51,r52,r53,r54,
r55,r56} for examples).

On the other hand, if the scaling attractor also has the property that its EoS of dark energy is 
smaller than $-1$, the second cosmological coincidence problem is alleviated at the same time. 
However, this is impossible in the interacting quintessence or k-essence scenario. 
Although the attractor's EoS is smaller than $-1$ in the interacting scalar phantom scenario, it is
impossible to cross the phantom divide $w_{de}=-1$, since the EoS of scalar phantom is always
smaller than $-1$. Fortunately, the EoS of vector-like dark energy can be smaller than $-1$,
while possessing a conventional positive kinetic term~\cite{r46}, in contrast to the scalar
phantom. Of course, the EoS of vector-like dark energy can be larger than $-1$ also.
Thus, crossing the phantom divide is possible in the vector-like dark energy model~\cite{r46}.
As is explicitly shown in this work, for suitable interaction forms [for instance, Cases~(II) and (IV)],
regardless of the model with inverse power-law or exponential potential, there are some attractors
with $w_A<-1$ while their corresponding $\Omega_A$ and $\Omega_m$ are comparable
in the interacting vector-like dark energy model. In the Case~(IV), {\em all} stable attractors
have these desirable properties. Even in the Case~(II), we can choose the model parameters to
avoid the attractor with $w_A>-1$, and the scaling attractor(s) with $w_A<-1$ becomes
the {\em unique} late time attractor(s). So, for a fairly wide range of initial conditions
with $w_A>-1$, the universe will eventually evolve to the scaling attractor(s) with $w_A<-1$.
Similar to the ordinary way to alleviate the (first) cosmological coincidence problem, the second
cosmological coincidence problem is alleviated at the same time.


\section{\label{sec7} Conclusion and discussion}
In summary, the cosmological evolution of the vector-like dark energy interacting with 
background perfect fluid is investigated in this work. We find that the first and second 
cosmological coincidence problems can be alleviated at the same time in this scenario. 
Our results obtained here may support the vector field to be one of the viable 
dark energy candidates. In particular, the feature of the vector-like dark energy that its EoS 
$w_A\equiv p_A/\rho_A$ can be smaller than $-1$ while possessing a conventional 
positive kinetic term is very attractive. While considering the interaction between the 
vector-like dark energy and background matter, the first and second cosmological 
coincidence problems can be alleviated at the same time. This is a profitable support to 
the vector-like dark energy. Of course, there are many remaining works to make this 
scenario more concrete, especially to fit the observational data to determine the realistic 
model parameters, which is beyond the main aim of the present work. 

The other issue is concerning the fate of our universe. It is easy to see that for all 
cases considered in this work, {\em all} stable attractors have $w_{eff}\geq -1$. Although 
the EoS of vector-like dark energy can be smaller than $-1$, the big rip never 
appears in this model. This is also in contrast to the ordinary phantom-like models.

Finally, we would like to mention that the gauge invariance is 
violated for the potential forms $V\left(A^{a2}\right)$ taken in 
this paper. Although in~\cite{r46}, it is 
argued that this kind of ``cosmic triad'' can naturally arise from 
gauge theory with a single $SU(2)$ gauge group because of 
the equations of motion can be written down in a gauge invariant 
form, one should be careful to this problem. 
For the cases of inverse power-law and exponential potential, 
however, it does no longer hold, which means a particular gauge 
has been taken.


\section*{ACKNOWLEDGMENTS}
We thank the anonymous referee for his/her quite useful comments 
and suggestions, which help us to deepen this work. H.W. is grateful 
to Zong-Kuan Guo, Yun-Song Piao, Bo Feng, Ding-Fang Zeng, Hui Li, 
Li-Ming Cao, Da-Wei Pang, Xin Zhang, Yi Zhang, Hua Bai, 
Hong-Sheng Zhang, Jia-Rui Sun, Xun Su, Wei-Shui Xu, Ya-Wen Sun, 
and Hao Ma for helpful discussions. This work is finished during R.G.C. 
visits the department of physics, Osaka university, supported by JSPS, the 
hospitality extended to him is appreciated. This work was supported 
in part by a grant from Chinese Academy of Sciences, and grants from 
NSFC, China (No. 10325525 and No. 90403029).


\section*{APPENDIX}
The particular parameter-space for the existence and stability of 
these critical points is considerably involved and verbose. Since 
our main aim here is just to point out the fact that it can exist 
and is stable, we do not present those very long expressions for 
the corresponding parameter-space. Of course, one can easily 
work out with the help of Mathematica. 

Instead of giving concrete expressions, we here give just some 
examples to support our statement. For the case of inverse 
power-law potential (Sec.~\ref{sec4}), Point~(P.II.3) of Case~(II) 
exists and is stable for parameters $\gamma=1$, $\beta=1/3$, and 
$n=2$, with eigenvalues $\{-0.866239,\, -1.36688-i\,3.29093,\, 
-1.36688+i\,3.29093\}$, while it is unstable for parameters 
$\gamma=1$, $\beta=1/3$, and $n=1$, with eigenvalues $\{-1.26235,\, 
2.13117-i\,2.04255,\, 2.13117+i\,2.04255\}$; Point~(P.IV.2) of 
Case~(IV) exists and is stable for parameters $\gamma=1$, 
$\sigma=1/2$, and $n=2$, with eigenvalues $\{-1.70481,\, 
-1.6976+i\,2.60705,\, -1.6976-i\,2.60705\}$, and for parameters 
$\gamma=1$, $\sigma=1/2$, and $n=3$, with eigenvalues $\{-1.1769,\, 
-2.32821+i\,2.93245,\, -2.32821-i\,2.93245\}$, while it is unstable 
for parameters $\gamma=1$, $\sigma=1/2$, and $n=1$, with 
eigenvalues $\{2.15106-i\,1.12309,\, -2.80212,\, 
2.15106+i\,1.12309\}$. Similarly, for the case of exponential 
potential (Sec.~\ref{sec5}), Point~(E.II.3) of Case~(II) exists 
and is stable for parameters $\gamma=1$, $\beta=1/3$, and 
$\lambda=6$, while Point~(E.II.4) is unstable for the same 
parameters; Point~(E.IV.2) of Case~(IV) exists and is stable for 
parameters $\gamma=1$, $\sigma=1/2$, and $\lambda=5$, while 
Point~(E.IV.3) is unstable for the same parameters. Note that in 
the above examples, we chose the demonstrative parameter 
$\gamma=1$ for dust matter, and $\beta=1/3$ or $\sigma=1/2$ to 
make $\Omega_m=1/3$, which are more realistic.

\renewcommand{\baselinestretch}{1.2}


\end{document}